\documentclass[aps,preprint,nofootinbib]{revtex4}
\usepackage{amssymb}
\usepackage{graphicx}
\usepackage{epsfig}
\usepackage[dvips]{color}

\begin{document}

\title{ Symmetry of Dirac Equation and Corresponding Phenomenology }

\author{
    Hong-Wei Ke$^{1}$, Zuo Li$^{2}$,  Jing-Ling Chen$^3$, Yi-Bing Ding$^4$ and
  Xue-Qian Li$^{2}$ }

\affiliation{
  $^{1}$ School of Science, Tianjin University, Tianjin 300072, China \\
  $^{2}$ School of Physics, Nankai University, Tianjin 300071,
  China\\
  $^{3}$ Theoretical Physics Division, Chern Institute of Mathematics,
Nankai University, Tianjin 300071, China,\\
  $^{4}$ Department of
Physics, Graduate University of Chinese Academy
  of Sciences, Beijing 100039, China}

\begin{abstract}
It has been suggested that the high symmetries in the Schr\"odinger
equation with the Coulomb or harmonic oscillator potentials may
remain in the corresponding relativistic Dirac equation. If the
principle is correct, in the Dirac equation  the potential should
have a form as ${(1+\beta)\over 2}V(r)$ where $V(r)$ is ${-e^2\over
r}$ for hydrogen atom and $\kappa r^2$ for harmonic oscillator.
However, in the case of hydrogen atom, by this combination the
spin-orbit coupling term would not exist and it is inconsistent with
the observational spectra of hydrogen atom, so that the symmetry of
SO(4) must reduce into SU(2). The governing mechanisms QED and QCD
which induce potential are vector-like theories, so at the leading
order only vector potential exists. However, the higher order
effects may cause a scalar fraction. In this work, we show that for
QED, the symmetry restoration is very small and some discussions on
the symmetry breaking are made. At the end, we briefly discuss the
QCD case and indicate that the situation for QCD is much more
complicated and interesting.
\end{abstract}

\maketitle

\section{introduction}

The Schr\"odinger equation is the basis of non-relativistic
quantum mechanics. It successfully explains the energy spectra of
hydrogen atom and harmonic oscillator, moreover as the orbit-spin
coupling is taken into account, the fine structure of hydrogen
atom is also calculated. All the predictions are consistent with
measurements of high precision.  In a complete theoretical
framework which respects the Lorentz covariance, the relativistic
Klein-Gorden equation for spin-zero particles and Dirac equation
for spin 1/2 particles substitute the Schr\"odinger equation to
describe physics. Surely, under the non-relativistic
approximation, they can reduce back to the Schr\"odinger equation
and meanwhile the orbit-spin coupling and relativistic correction
terms which were added into the non-relativistic Hamiltonian by
hand, automatically appear. Nowadays, as the detection technique
and instrumentation are ceaselessly improved, small deviation
between theoretical predictions and observational data may be
exposed, such as the Lamb shift\cite{Lamb,Lamb2,Lamb3}, and it
enables one to investigate the tiny difference caused by various
theoretical models.

No wonder, as is well known, symmetry is extremely important in
all fields of physics. Generally, the Schr\"odinger equation for
hydrogen atom which is of a Coulomb-type potential of N dimensions
possesses an SO(N+1) symmetry. It is not a geometric symmetry, but
a dynamical one. The SO(4) group in three dimensions can be
decomposed into a direct product of two independent SU(2) groups
which correspond to spin and orbital angular momenta respectively
\cite{Greiner}. Recently, Zhang, Fu and Chen\cite{Chen} indicate
that the Schr\"odinger equation with a central force fields has an
explicit SO(4) symmetry, but generally the Dirac equation with the
same potential does not possess the SO(4) symmetry. The harmonic
oscillator has the similar situation. The Schr\"odinger equation
has an SU(3) symmetry, but the corresponding Dirac equation does
not. Ginocchio\cite{Ginocchio,Ginocchio2} investigated this
problem and obtained a modified Dirac equation which possesses the
SU(3) symmetry and can also reduce into the original Schr\"odinger
equation under non-relativistic approximation.

In the central force field the non-relativistic Hamiltonian is
$H=\frac{p^2}{2m}+V(r)$, turning into the operator form, one only
needs to replace ${\bf p}$ into $-i\hbar\nabla$ and does not change
the potential $V(r)$ in the configuration representation, the
Schr\"odinger equation of the central force would retain the
symmetry of $V(r)$ (${-e^2\over r}$ for hydrogen and $\kappa r^2$
for harmonic oscillator). By contrast, in the relativistic theory,
the Hamiltonian of a free particle should be $H=\sqrt{p^2+m^2}$,
thus if one writes it into an operator form, the square root
$\sqrt{p^2+m^2}$ needs to be replaced by $\mathbf{\alpha}\cdot
\mathbf{p}+m\beta$. However, if there was a potential $V(r)$ in the
non-relativistic Hamiltonian, then how to write a corresponding term
in the Hamiltonian when the Dirac equation replaces the
Schr\"odinger equation, would be a question. In other words, one can
ask whether he needs to change its form. The authors of
Ref.\cite{Chen} showed that if we do not make a proper changes on
the form of $V(r)$, the corresponding Dirac equation with the
central force filed fails to possess the original SO(4) symmetry at
all. Alternatively, if we re-write it as $V(r)(1+\beta)/2$, the
SO(4) symmetry would be maintained. Generally, The non-relativistic
theory is a natural approximation of the relativistic theory, thus
one may suppose that the symmetry degree in the relativistic theory
should be higher than the non-relativistic one. Therefore, there is
a compelling reason to believe that the Dirac equation of hydrogen
atom should respect the SO(4) symmetry.

However, the physics does not promise the symmetry argument. As the
authors of Ref.\cite{Chen} indicate, when the Hamiltonian in the
Dirac equation is of the form : $\alpha\cdot {\bf p}+\beta M+{1\over
2}(1+\beta)V$, as it reduces into the Schr\"odinger equation, the
spin-orbit coupling term does not exist, namely the $2P_{1/2}$ and
$2P_{3/2}$ states should be degenerate, i.e. the fine structure of
hydrogen would not be experimentally observed. That conclusion is
definitely contrary to the experimental measurement.

Let us investigate the origin of the scalar and vector potentials in
the Dirac equations based on the quantum field theory (QFT).
Considering all possible Lorentz structures of the interactions
which conserve parity, there are five types of interactions between
two charged fermions \cite{Lucha}: $1\otimes 1\sim V_s(r)$;
$\gamma_5\otimes \gamma_5\sim 0$; $\gamma^{\mu}\otimes
\gamma_{\mu}\sim V_V(r)$;
$\gamma^{\mu}\gamma_5\otimes\gamma_{\mu}\gamma_5\sim {\bf S}_1\cdot
{\bf S}_2V_A(r)$ and $\sigma_{\mu\nu}\otimes \sigma^{\mu\nu}\sim
{\bf S}_1\cdot {\bf S}_2V_T(r)$. As is believed, the underlying
theory which is responsible for electromagnetic processes is QED
which is a vector-like gauge theory.  At the leading order only
$\gamma^{\mu}\otimes\gamma_{\mu}$ exists, which induces merely the
vector potential $V_V(r)$ by exchanging a virtual photon. However,
the higher order effects, i.e. multi-photon exchanges can cause
scalar and tensor interactions. If we ignore the part $V_A(r)$ and
$V_T(r)$ which are related to the hyper-fine structure in hydrogen
atom case and tiny, only the scalar potential should emerge.

The observation raises a serious problem that the high symmetry in
the Schr\"odinger equation which is a non-relativistic approximation
of the Dirac equation might be superficial. When the spin-orbit
coupling is taken into account, the symmetry would disappear. In
other words, in the Schr\"odinger equation with only the Coulomb's
potential ${-e^2\over r}$, the apparent SO(4) symmetry does not
survive in the relativistic Dirac equation. Interesting issue is how
much the symmetry can be restored. In fact, in some literatures
which is aiming to study phenomenology, one writes the potential as
$(b+a\beta)V(r)$ with $a+b=1$, and lets $a$ be a free parameter to
be fixed by experimental data.

With the form provided in Ref.
\cite{SVPotential,SVPotential2,SVPotential3,SVPotential4,SVPotential5,SVPotential6}:
$H=\alpha\cdot {\bf p}+\beta(M+V_s)+V_V$ and by our above
discussion, we write it as $H=\alpha\cdot {\bf p}+\beta
M+(b+a\beta)V$, then following Greiner's calculation which
includes all possible combinations of $a $ and $b$, we can use the
hydrogen atom spectrum which has been measured with high precision
to determine a and b and see to what degree the SO(4) symmetry can
be restored or never. We find that for the hydrogen atom, the
fraction of the scalar potential is only at order of 0.1\%, namely
in the QED case (responsible for the hydrogen atom) the symmetry
breaking is almost complete, i.e. $SO(4)\longrightarrow SU(2)$.

Based on the quantum field theory, the global and local symmetries
which reside in the Lagrangian are real, instead, the symmetries in
the Schr\"odinger equation are indeed superficial. On other aspect,
the higher effects can partly restore the symmetry in the potential,
in fact, the information can help us understanding the effects of
higher orders in QED and even in QCD where higher orders and
non-perturbative  effects may play a crucial role.

After this long introduction, we will present our numerical results
on the hydrogen atom spectrum and determine the fraction of scalar
potential in section II. Then we analyze the origin of the scalar
potential based on the QED theory and discuss the breakdown of
SO(4). The last section is devoted to our conclusion and meanwhile
we briefly discuss the QCD case and indicate that for QCD the
situation is much more complicated and the symmetry analysis might
help us to get a better understanding of the non-perturbative QCD
effects and higher order contributions.

\section{Dirac equation for hydrogen atom}

In order to further explore the symmetry problem, we investigate the
symmetry issue of the most familiar hydrogen atom which has been
thoroughly studied from both theoretical and experimental aspects.
In the case of hydrogen atom, the interaction between the electron
and proton is purely electromagnetic, so that one can determine to
what degree the SO(4) symmetry is broken.

The Dirac equation with an unbroken SO(4) symmetry has the form
\cite{Chen}
\begin{eqnarray}\label{1s1}
[c\hat{\alpha} \cdot
\hat{p}+mc^2\beta+\frac{1+\beta}{2}V(r)-E]\psi=0,
\end{eqnarray}
and for hydrogen atom $V(r)=-\frac{e^2}{r}$.

In this expression the vector and scalar potentials coexist with the
same weight. Its solution is explicitly given in Ref.\cite{Greiner},
\begin{eqnarray}\label{1s2}
[c\hat{\alpha} \cdot \hat{p}+\beta (mc^2+V_2)-(E-V_1)]\psi=0,
\end{eqnarray}
where  $V_1$ and $V_2$ are vector and scalar potentials
respectively. If one chooses $V_1=V_2=-\hbar
c\frac{\alpha}{2r}=-\frac{e^2}{2r},$ he can obtain Eq.\ref{1s1}. The
spectrum is
\begin{eqnarray}\label{1s3}
E=m[1-\frac{2\alpha^2}{\alpha^2+4n^2}],
\end{eqnarray}
where the natural unit system $c=\hbar=1$ is adopted and the
principal quantum number is $n=n'+j+\frac{1}{2}$.

From Eq.(\ref{1s3}) we can see that the energy spectrum is
determined only by the principal quantum $n$ i.e. the states with
various $j-$values are degenerate and there would be no
fine-structures, and it is the picture described by the
Schr\"odinger equation whose potential does not involve the
spin-orbit couplings.

Generally the Dirac equation for hydrogen atom as only the vector
potential being involved, takes the form which is given in
ordinary textbooks of quantum mechanics:
\begin{eqnarray}\label{2s4}
[c\hat{\alpha} \cdot \hat{p}+mc^2\beta-{e^2\over r}-E]\psi=0.
\end{eqnarray}

The corresponding eigen-energies are
\begin{eqnarray}\label{2s5}
&&E=m[1+\frac{\alpha^2}{(s+n')^2}]^{-\frac{1}{2}},
\end{eqnarray}
where $s=\sqrt{(j+\frac{1}{2})^2-\alpha^2}$ and the energies are
explicitly related to quantum number $j$.

We present the eigen-energies corresponding to various $j$ and $l$
when $n=1,2$ in the following table,
\begin{table}[!h]
\caption{}
\begin{center}    \label{t11}
\begin{tabular}{|c|c|c|c|c|c|c|c|}
  \hline
 $n$ & $n'$ &$j$ &$K$ &$l$ & structure &$E_1$(eV)& $E_2$(eV)\\\hline
 $1$& 0& $\frac{1}{2}$ &1 &0 & $1s_{\frac{1}{2}}$&13.605433412 &13.605071149\\\hline
  $2$& 0& $\frac{3}{2}$ &2 &1 & $2p_{\frac{3}{2}}$&3.400994390 &3.400971749\\\hline
  $2$& 1& $\frac{1}{2}$ &-1 &1 & $2p_{\frac{1}{2}}$&3.401039674&3.400971749\\\hline
  $2$& 1& $\frac{1}{2}$ &1 &0 & $2s_{\frac{1}{2}}$&3.401039674&3.400971749\\\hline
\end{tabular}
\end{center}
\end{table}
where $E_1$ and $E_2$ are calculated in terms of Eqs.(\ref{2s5}) and
(\ref{1s3}) respectively and the mass of electron and fine-structure
constant are set according to Ref. \cite{PDG06}. Since the spectrum
of hydrogen has been experimentally measured with very high
accuracy, we keep more significant figures in our calculation
results accordingly. It is shown later that some small distinctions
between theory and data only appear at the  last few digits of the
number.

{\section{Symmetry is restored partly or never}

As discussed above, let us rewrite the Dirac equation as
\begin{eqnarray}\label{3s1}
[c\hat{\alpha} \cdot \hat{p}+mc^2\beta+bV(r)+a\beta V(r)-E]\psi=0,
\end{eqnarray}
with $a$ constraint $a+b=1$ which ensures this equation to reduce
back to the familiar Schr\"odinger equation, but now we suppose
$a$ and $b$ to be free parameters (in fact, only $a$ or $b$ is
free by the constraint).

It is easy to obtain the eigen-energy of hydrogen according to Ref.
\cite{Greiner} as
\begin{eqnarray}\label{3s2}
E &=&
m\{\frac{-(1-a)a\alpha^2}{(1-a)^2\alpha^2+(n-j-\frac{1}{2}+s)^2}+[(\frac{(1-a)a\alpha^2}{(1-a)^2\alpha^2+(n-j-\frac{1}{2}+s)^2})^2
\nonumber\\
&&-\frac{a^2\alpha^2-(n-j-\frac{1}{2}+s)^2}{(1-a)^2\alpha^2+(n-j-\frac{1}{2}+s)^2}]^{\frac{1}{2}}\},
\end{eqnarray}

Generally, one can determine the parameter $a$ by fitting the data
of hydrogen atom. It is noted that all the measured data of
hydrogen atoms involve the effects of not only fine structure
which results in the splitting of energies corresponding to
various j-values, but also the hyper-fine structure and Lamb
shift. The hyper-fine structure is due to the interaction between
the spins of nucleus (proton in the hydrogen case) and electron,
while the Lamb shift is caused by the virtual photon effects. To
extract the necessary information about the fine structure, one
can either carefully carry out a theoretical estimate to get rid
of the influences from other effects or choose special transition
modes where only the physical mechanism we concern plays a
dominant role. In fact, we may ignore the effects of the hyperfine
structure, because it is suppressed by a factor $m_e/m_p\sim
1/2000$ compared to the fine structure caused by the spin-orbit
coupling \cite{Landau,Landau2}.

The energy gap between $2p_{\frac{3}{2}}$ and $2p_{\frac{1}{2}}$
may be a good candidate to study the splitting induced by
spin-orbit coupling, since they are in $p$ wave and the Lamb shift
is small\cite{2pLamb}. The level-crossing value between
$2p_{\frac{3}{2}}$ and $2p_{\frac{1}{2}}$ has been measured as
$10969.20\pm0.06$MHz\cite{2p2p}. Using this value we obtain
$a=-(8.93\pm0.02)\times 10^{-4}$ ( $a/b=-0.000893$) which
manifests the restoration degree of the SO(4) symmetry.

There are three comments,

(1) Since $a$ is a very small number, Eq.(\ref{3s1})is very close
to Eq.(\ref{2s4}), i.e. the symmetry breaking in the QED case is
almost complete.

(2) Especially, $a$ is negative. Since this value is obtained by
fitting data, we should suppose that it is valid. Question is what
the negative sign implies? As we discuss above, the SO(4) symmetry
demands $a=b=1/2$, if $a$ were a small positive value, we could
allege that the SO(4) symmetry were partly restored, but now it is a
small negative value, what conclusion can we draw? We will come to
this question at the last section.

(3) Even though there is a scalar potential in Eq.(\ref{3s1}), the
mechanisms which induce the Lamb shift and hyper-fine-structure
are not involved in the Dirac equation (Eq.(\ref{3s1})) as
aforementioned.

\begin{table}[!h]
\caption{}
\begin{center}    \label{t22}
\begin{tabular}{|c|c|c|c|c|c|c|c|}
  \hline
 $n$ & $n'$ &$j$ &$K$ &$l$ & spectrum & eigen-values (eV)\\\hline
 $1$& 0& $\frac{1}{2}$ &1 &0 & $1s_{\frac{1}{2}}$&13.605434057 \\\hline
  $2$& 0& $\frac{3}{2}$ &2 &1 & $2p_{\frac{3}{2}}$&3.400994431 \\\hline
  $2$& 1& $\frac{1}{2}$ &-1 &1 & $2p_{\frac{1}{2}}$&3.401039795\\\hline
  $2$& 1& $\frac{1}{2}$ &1 &0 & $2s_{\frac{1}{2}}$&3.401039795\\\hline
\end{tabular}
\end{center}
\end{table}

\section{Origin of scalar potential in quantum field theory}

In principle, we can make a general discussion about how the scalar
potential emerges in the relativistic quantum mechanics. The least
action principle in quantum mechanics tells us how to construct the
Hamiltonian for a charged particle in electromagnetic field by
transforming operator ${\bf p}$ into ${\bf p}-e{\bf A}$ and $E$ into
$E-e\phi(r)$ where $\phi(r)$ is the zeroth component of
$A^{\mu}$\footnote{Here to avoid unnecessary ambiguity, we do not
use the terminology commonly used in electrodynamics: scalar and
vector potentials which obviously have different meaning from what
we refer in this work. }. The corresponding Dirac equation would be
of the form $[\mathbf{\alpha}\cdot
\mathbf{p}+m\beta+\phi(r)]\psi=E\psi$ where only the vector
potential exists, because the energy $E$ has the same Lorentz
structure as $\phi$ and it is exactly $V_V(r)$ we concern. If the
non-relativistic Hamilton can be written as
$H=\sqrt{p^2+(m+V(r))^2}$, after the transformation, we can obtain a
different equation $[\mathbf{\alpha}\cdot
\mathbf{p}+(m+V(r))\beta]\psi=E\psi$, because now the potential $V$
has the same Lorentz structure as the mass and then $V$ is a scalar
potential. However there is no  reason to write the Hamiltonian in
that way.

Now let us turn to the QED theory which is viewed as the correct
theory to be responsible for all the electromagnetic processes, at
the quantum field theory level. As aforementioned, the QED is a
vector coupling theory, so that at the leading order the scalar
potential does not exist. One can ask if the scalar potential
emerges at higher orders. The answer is positive: actually we can
deduce a scalar coupling from high order QED. Due to the fundamental
coupling between electron and photon
$\overline\psi_e\gamma_{\mu}\psi_eA^{\mu}$, we achieve the vector
potential from a single photon exchange\cite{Greiner,Zuber}.

\begin{figure}
\begin{center}
\begin{tabular}{cc}
\includegraphics[width=5cm]{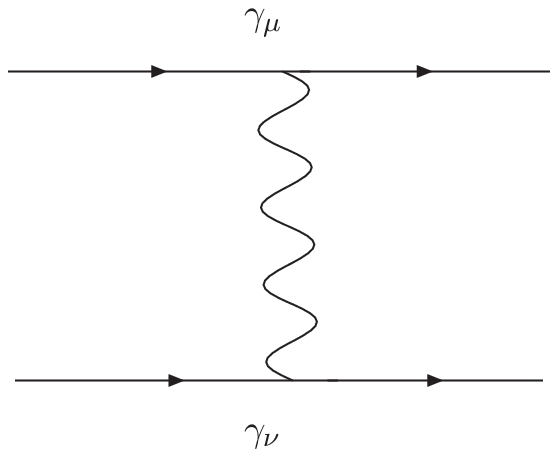}
\end{tabular}
\end{center}
\caption{ } \label{fig:fbeta}
\end{figure}

When the high order effects are taken into account, namely, as we
consider the loop contributions, the scalar coupling appears. The
next to leading order Feynman diagrams with two photons exchanged
between electrons are shown in Fig.2 and 3. Omitting irrelevant
factors in the amplitude, we can write it as
 \begin{eqnarray}\label{4s1}
M\propto \int
\bar{U}(P_b)\gamma_\rho\frac{P_1\!\!\slash+m_1}{P_1^2-m_1^2}\gamma_\mu
U(P_a){d^4P_1\over (2\pi)^4}
\bar{U}(P_d)\gamma_\omega\frac{P_2\!\!\!\slash+m_2}{P_2^2-m_2^2}\gamma_\nu
U(P_c)g^{\rho\omega}g^{\mu\nu}\frac{1}{(P_a-P_1)^2(P_1-P_b)^2}.
\end{eqnarray}

  \begin{figure}
\begin{center}
\begin{tabular}{cc}
\includegraphics[width=7cm]{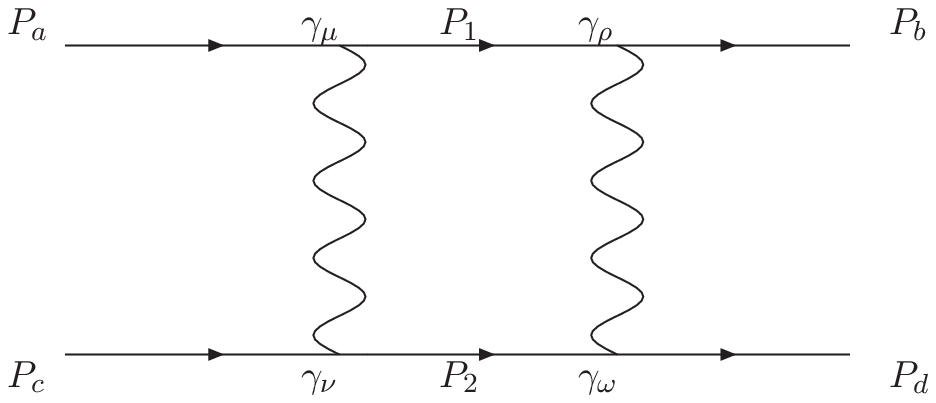}
\end{tabular}
\end{center}
\caption{ } \label{fig:fbeta}
\end{figure}

     \begin{figure}
\begin{center}
\begin{tabular}{cc}
\includegraphics[width=7cm]{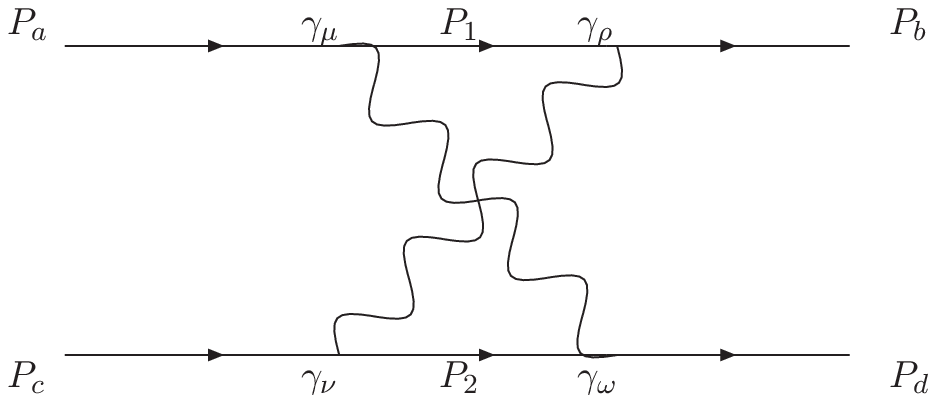}
\end{tabular}
\end{center}
\caption{ } \label{fig:fbeta}
\end{figure}
The term
  \begin{eqnarray}\label{4s2}
\bar{U}(P_b)\gamma_\rho\frac{-m_1}{P_1^2-m_1^2}\gamma_\mu U(P_a)
\bar{U}(P_d)\gamma_\omega\frac{-m_2}{P_2^2-m_2^2}\gamma_\nu
U(P_c)g^{\rho\omega}g^{\mu\nu}\frac{1}{(P_a-P_1)^2(P_1-P_b)^2},
\end{eqnarray}
can be rewritten as
  \begin{eqnarray}\label{4s3}
\frac{m_1m_2}{(P_1^2-m_1^2)(P_2^2-m_2^2)}\bar{U}(P_b)\gamma_\rho\gamma_\mu
U(P_a) \bar{U}(P_d)\gamma_\omega\gamma_\nu
U(P_c)g^{\rho\omega}g^{\mu\nu}\frac{1}{(P_a-P_1)^2(P_1-P_b)^2}.
\end{eqnarray}

A simple relation
  \begin{eqnarray}\label{4s4}
\gamma_\alpha\gamma_\beta=\frac{1}{2}(\gamma_\alpha\gamma_\beta-\gamma_\beta\gamma_\alpha)+
\frac{1}{2}(\gamma_\alpha\gamma_\beta+\gamma_\beta\gamma_\alpha)=g_{\alpha\beta}-i\sigma_{\alpha\beta},
\end{eqnarray}
leads to
  \begin{eqnarray}\label{4s3}
&&\frac{m_1m_2}{(P_1^2-m_1^2)(P_2^2-m_2^2)}\bar{U}(P_b)g_{\rho\mu}
U(P_a) \bar{U}(P_d)g_{\omega\nu}
U(P_c)g^{\rho\omega}g^{\mu\nu}\nonumber\\&&=\frac{m_1m_2}{(P_1^2+m_1^2)(P_2^2+m_2^2)}\bar{U}(P_b)
U(P_a) \bar{U}(P_d) U(P_c)\frac{1}{(P_a-P_1)^2(P_1-P_b)^2},
\end{eqnarray}
which is explicitly a $1\times 1$ type coupling and eventually
induces a scalar potential. The Feynman diagram in Fig.3 also
results in similar scalar coupling.

Now let us turn to study the form of the induced potential. In fact,
the integration is complicated and usually cannot be analytically
derived. This is similar to deriving a potential between two
nucleons induced by two-pion exchange. Instead of making a full
integration over $q$ which is supposed to correspond to the one-loop
radiative correction to the scattering amplitude, we would like to
derive the main piece of an effective potential between two charged
particles. Let us focus on the two photon propagators in
Eq.(\ref{4s3}), we write the term as
\begin{equation}\label{sub}
{1\over q^2(q+P)^2},
\end{equation}
where $q=P_1-P_b$ and $P=P_b-P_a$, and the two fermion propagators
are quenched.

The standard way\cite{Landau,Landau2} of obtaining a potential in
QFT is to carry out a Fourier transformation on ${\bf P}$ in the
transition amplitude. Now let us see the integration

\begin{eqnarray}\label{intp1}
I&=&\int\frac{dq^4}{(2\pi)^4} \int {e^{i{\bf P}\cdot{\bf r}}\over q^2[(q_0+P_0)^2-(\mathbf{q}+\mathbf{P})^2]}d^3\mathbf{P},\nonumber\\
&=&\int\frac{dq^4}{(2\pi)^4} \int {e^{i{\bf P}\cdot{\bf r}}\over
q^2[q_0^2-(\mathbf{q}+\mathbf{P})^2]}d^3\mathbf{P},
\end{eqnarray}
where the loop integration over four-momentum $q$ can be carried out
in the regular way. For the Fourier integration, following the
standard, we set $ P_0=0$. Shifting the integration variable ${\bf
P}$ into $\bf P'=P+q$, we have
\begin{eqnarray}\label{intp2}
I&=&\int\frac{dq^4}{(2\pi)^4} \int {e^{i({\bf P'-q})\cdot{\bf
r}}\over q^2[q_0^2-\bf P'^2]}d^3\mathbf{P'}\nonumber\\
&=&\int\frac{dq^4}{(2\pi)^4}{e^{-i{\bf q}\cdot{\bf r}}\over q^2}
\int {e^{i({\bf P'})\cdot{\bf r}}\over [q_0^2-\bf
P'^2]}d^3\mathbf{P'}.
\end{eqnarray}
The contour integration is standard and one immediately obtains the
results in terms of the residue principle as
\begin{eqnarray}\label{Coulomb}
I&=&\int\frac{dq^4}{(2\pi)^4}{e^{-i{\bf q}\cdot{\bf r}}\over q^2}
\frac{e^{-|q_0||\bf
r|}}{\mathbf{r}}\nonumber\\&=&\frac{1}{\mathbf{r}}\int\frac{dq^4}{(2\pi)^4}{e^{-i{\bf
q}\cdot{\bf r}}\over q^2} e^{-|q_0||\bf r|}.
\end{eqnarray}

It shows that the leading term in the induced potential is a
generalized Coulomb-type, even though the numerator in
Eq.(\ref{Coulomb}) is not a simple  rational number, but an integration. It is worth noting that the
numerator in Eq.(\ref{Coulomb}) corresponds to the loop integration
which is not Lorentz invariant because we have already chosen the
rest frame of the the two-fermion system  as our reference frame. To
really calculate it, one may need to deal with possible infrared
divergence which appears for soft photon propagation, but anyhow, it
is a number. It is beyond the scope of this work, here we only show
the characteristic of the potential induced by the two-photon
exchange. Indeed, such a potential induced by the two-photon
exchange is very complicated and it is in analog to the interaction
between nucleons induced by two-pion exchange which has been
thoroughly studied by many authors\cite{Chemtob}. The potential form
is not easy to be framed, generally can be better described by the
Bethe-Salpeter equation \cite{Rocha}. In Ref. \cite{Rocha}, the
authors gave a very complicated potential form, but one can notice
that the leading piece is still the Yukawa-type one which
corresponds to the Coulomb-type in the case of two photon-exchange.

As carrying out above theoretical calculations, we only consider the
scalar Coulomb type potential in Eq.(\ref{4s4}), but definitely when
fitting data,   contributions from all other pieces are involved
altogether. It seems that our theoretical calculation is not
complete. However, the contributions from other pieces are either of
the same order of magnitude or, most probably, smaller than the
contributions of the Coulomb piece. Therefore, as we discuss the
symmetry breaking and restoration in this work, we only concern the
order of magnitude, thus our results which only involve contribution
from the Coulomb piece do make sense. In other words, our derivation
is not rigorous, but is illustrative and persuasive. We will carry
out a complete derivation of the two-photon induced potential in our
later work. It also interesting to note that the sign of the
Coulomb-type potential induced by two-photon exchange is opposite to
that induced by one-photon exchange. It can naturally explain the
minus sign of $a$ obtained by fitting data.

Because the scaler coupling  arise at the next to leading order so
the scalar potential should generally undergo a loop suppression. By
a simple analysis of order of magnitude, this suppression factor is
proportional to $\alpha/\pi=\frac{1}{137\pi}\simeq 0.00232$ relative
to the vector potential. Definitely the concrete suppression factor
should be derived by explicitly calculating the loop diagrams and
numerically evaluate the contributions. But as the order
$O(\alpha^2)$ and even higher orders can contribute to the scalar
potential, we would like to obtain the suppression factor $a/b$
phenomenologically. The number on $a/b$ which is obtained by fitting
the spectrum of hydrogen and the value was given in last section,
one can notice that it is consistent with our qualitative analysis.

\section{Conclusion and discussion}

It is a well-known fact that symmetry plays an important role in
physics. Not only the  geometric symmetries are obviously
important, but also  dynamical symmetries are fundamental and
crucial for determining the physics. For example, It has been a
long time, people notice that there is an extra symmetry $[{\bf
R}, H]=0$ with ${\bf R}={1\over 2m}({\bf p}\times {\bf L}-{\bf
L}\times{\bf p})-{e^2\over r}\mathbf{\hat r}$ for the hydrogen
atom. This symmetry finally leads to the $n^2$ degeneracy of the
hydrogen levels \cite{Baym}. The symmetry corresponds to two
``angular momenta" and the generators are respectively
$${\bf M}={{\bf L}+{\bf K}\over 2};$$
and
$${\bf N}={{\bf L}-{\bf K}\over 2},$$
where ${\bf K}\equiv \sqrt{{-m\over 2H}}{\bf R}$. The two ``angular
momenta" symmetries are involved in a larger group SO(4).

Similarly, it is indicated that the Hamiltonian of harmonic
oscillator has an SU(3) dynamical symmetry. Since the
Schr\"odinger equation is the relativistic approximation of the
relativistic Dirac equation for fermions, one can ask if the
larger symmetries are maintained in the corresponding Dirac
equations? The symmetry causes degeneracy and the symmetry
breaking leads to splitting of energy levels, thus to answer the
question, we need to study the observational energy level
splitting. Ginocchio \cite{Ginocchio,Ginocchio2} and Zhang et al.
\cite{Chen} indicate that to maintain the higher symmetry in the
Dirac equations for harmonic oscillator and hydrogen atom, the
scalar and vector potentials must have the same weight, i.e. the
combination is of the form $(1+\beta)V(r)$. If one changes the
combination into $(a+b\beta)V(r)$ with $a+b=1$, as he reduces the
Dirac equation into  the non-relativistic Schr\"odinger equation,
at the leading order, the combination does not induce distinction
because $\beta=\left(\begin{array}{cc} I & 0\\ 0 & -I
\end{array}\right)$, only the left corner entry plays a role. But to the
next order, the right corner entry joins the game and $a$ and $b$
terms would have different contributions and the superficial
symmetry is broken unless $a=b=1/2$. If so, the spin-orbit
coupling does not exist and fine structure (energy splitting)
disappears. It obviously contradicts to the experimental
observation.

The underlying theory is QED which is a vector-like theory, and
its leading order contribution to the potential is via the
one-photon exchange which definitely results in a vector potential
in the Dirac equation. Even though the higher order QED effects
would induce the scalar potential, it should be much suppressed
comparing to the leading term. The data tell us that the fraction
of the scalar potential is at order of 0.1\% which is consistent
with our estimation based on the quantum field theory. We obtain
the value of $a$  as $a=-(8.93\pm 0.02)$ ( $a/b=-0.00089$) by
fitting the hydrogen spectrum, and it seems that the value further
deviates from $a=b=1/2$ which is demanded by the SO(4) symmetry.
Does the negative sign really mean that the symmetry is not partly
restored? Since the value of $a$ is too small, some theoretical
and experimental uncertainties can cause a small fluctuation in
the value, so that we cannot yet draw a definite conclusion that
the symmetry is not restored. To confirm its sign, one must pursue
this study in a case where the restoration might be sizable, and
no doubt it is the QCD case.

Since the higher order contributions of QCD would be much more
important because $\alpha_s$ is large at lower energies, so one
can expect that in the QCD case, such symmetry restoration might
be significant. Castro and
Franklin\cite{Franklin,Franklin2,Franklin3,Franklin4} have worked
out the Dirac equation where the authors considered an arbitrary
combination of the scalar and vector potentials for both the
Coulomb and linear pieces and obtain the wavefunction of ground
state of a $q\bar q$ bound state. Interesting aspect is that they
find that the dominant part in the linear confinement potential is
the scalar part. That conclusion is obviously different from the
QED case. Since the linear potential is fully due to the
long-distance effects of QCD and may be caused by the
non-perturbative contributions, we cannot derive it from QFT so
far (in the the lattice theory, the area theorem may imply a
linear potential form, but cannot distinguish the vector and
scalar fractions at all.), the argument about the origin of scalar
potential given in last section becomes incomplete. We will
further pursue this interesting issue and it is the goal of our
next work\cite{Ke}.

Our conclusion is whether the high symmetries in the Schr\"odiner
equation may exist in the corresponding Dirac equation needs careful
studies. In the case of hydrogen atom, it definitely does not. We
show that the symmetry breaking is almost complete, from the
experimental and theoretical aspects.

Indeed it would an interesting problem because we explicitly show
the breaking for hydrogen atom, but for other kinds of potentials
which may be induced by various mechanisms, especially the
phenomenological potentials such as the Cornell-type potential which
is responsible for the quark confinement, whether such symmetry can
be kept in the Dirac equation is worth further investigation.

\section*{Acknowledgments}
This work is supported in part by the National Natural Science
Foundation of China and the Special Grant for Ph.D programs of the
Education Ministry of China, one of us (Ke) would like to thank
the Tianjin University for financial support.

\end{document}